\documentclass[aps,prb,twocolumn,showpacs,superscriptaddress,amsmath,amssymb,longbibliography]{revtex4-1}

\usepackage[pdftex]{graphicx}
\usepackage[pdftex,bookmarks=true,bookmarksopen,bookmarksnumbered,
                colorlinks,
                linkcolor=blue,
                citecolor=blue]{hyperref}
\usepackage{dcolumn}
\usepackage{bm}
\usepackage{braket,amsmath}

\begin{document}


\title{Breaking the rotating wave approximation for a strongly-driven, dressed, single electron spin}

\author{Arne Laucht}
\email{a.laucht@unsw.edu.au}
\affiliation{Centre for Quantum Computation and Communication Technology, School of Electrical Engineering \& Telecommunications, UNSW Australia, Sydney, New South Wales 2052, Australia}
\author{Stephanie Simmons}
\altaffiliation[Now at ]{Physics Department, Simon Fraser University, British Columbia, Canada}
\affiliation{Centre for Quantum Computation and Communication Technology, School of Electrical Engineering \& Telecommunications, UNSW Australia, Sydney, New South Wales 2052, Australia}
\author{Rachpon Kalra}
\altaffiliation[Now at ]{Queensland Quantum Optics Laboratory, University of Queensland, Brisbane, Queensland 4072, Australia}
\affiliation{Centre for Quantum Computation and Communication Technology, School of Electrical Engineering \& Telecommunications, UNSW Australia, Sydney, New South Wales 2052, Australia}
\author{Guilherme Tosi}
\affiliation{Centre for Quantum Computation and Communication Technology, School of Electrical Engineering \& Telecommunications, UNSW Australia, Sydney, New South Wales 2052, Australia}
\author{Juan P. Dehollain}
\altaffiliation[Now at ]{QuTech \& Kavli Institute of Nanoscience, TU Delft, 2628CJ Delft, The Netherlands}
\affiliation{Centre for Quantum Computation and Communication Technology, School of Electrical Engineering \& Telecommunications, UNSW Australia, Sydney, New South Wales 2052, Australia}
\author{Juha T. Muhonen}
\altaffiliation[Now at ]{Center for Nanophotonics, FOM Institute AMOLF, Science Park 104, 1098 XG, Amsterdam, The Netherlands}
\affiliation{Centre for Quantum Computation and Communication Technology, School of Electrical Engineering \& Telecommunications, UNSW Australia, Sydney, New South Wales 2052, Australia}
\author{Solomon Freer}
\affiliation{Centre for Quantum Computation and Communication Technology, School of Electrical Engineering \& Telecommunications, UNSW Australia, Sydney, New South Wales 2052, Australia}
\author{Fay E. Hudson}
\affiliation{Centre for Quantum Computation and Communication Technology, School of Electrical Engineering \& Telecommunications, UNSW Australia, Sydney, New South Wales 2052, Australia}
\author{Kohei M. Itoh}
\affiliation{School of Fundamental Science and Technology, Keio University, 3-14-1 Hiyoshi, 223-8522, Japan}
\author{David N. Jamieson}
\affiliation{Centre for Quantum Computation and Communication Technology, School of Physics, University of Melbourne,
Melbourne, Victoria 3010, Australia}
\author{Jeffrey C. McCallum}
\affiliation{Centre for Quantum Computation and Communication Technology, School of Physics, University of Melbourne,
Melbourne, Victoria 3010, Australia}
\author{Andrew S. Dzurak}
\affiliation{Centre for Quantum Computation and Communication Technology, School of Electrical Engineering \& Telecommunications, UNSW Australia, Sydney, New South Wales 2052, Australia}
\author{Andrea Morello}
\affiliation{Centre for Quantum Computation and Communication Technology, School of Electrical Engineering \& Telecommunications, UNSW Australia, Sydney, New South Wales 2052, Australia}

\date{\today}

\begin{abstract}
We investigate the dynamics of a strongly-driven, microwave-dressed, donor-bound electron spin qubit in silicon. A resonant oscillating magnetic field $B_1$ is used to dress the electron spin and create a new quantum system with a level splitting proportional to $B_1$. The dressed two-level system can then be driven by modulating the detuning $\Delta\nu$ between the microwave source frequency $\nu_{\rm MW}$ and the electron spin transition frequency $\nu_e$ at the frequency of the level splitting. The resulting dressed qubit Rabi frequency $\Omega_{R\rho}$ is defined by the modulation amplitude, which can be made comparable to the level splitting using frequency modulation on the microwave source. This allows us to investigate the regime where the rotating wave approximation breaks down, without requiring microwave power levels that would be incompatible with a cryogenic environment. We observe clear deviations from normal Rabi oscillations and can numerically simulate the time evolution of the states in excellent agreement with the experimental data. 
\end{abstract}


\maketitle
Spins in semiconductors are excellent candidates for quantum bits with long coherence times, outstanding controllability and demonstrated two-qubit logic gates~\cite{Petta2005,Koppens2006,Maune2012,Pla2012,Pla2013,Muhonen2014,Veldhorst2014,Veldhorst2015}. Furthermore, they are the key ingredient to some of the most promising solid-state architectures for quantum computation~\cite{Kane1998,Loss1998,Taylor2005,Hollenberg2006,Hill2015,Tosi2015}. In these architectures, controlled gate operations are, ideally, performed by placing the spins in a static magnetic field $B_0$ and then applying a perpendicular magnetic field $B_1$ that rotates at a frequency equal to the spin Lamor frequency $\nu_{\rm L}$~\cite{Cohen1992}. However, in most cases the time-dependent magnetic field is linearly polarized, i.e. oscillating and not rotating. An oscillating field can be decomposed into the vector sum of two fields rotating in opposite directions, of which only one coincides with the natural direction of the spin's Larmor precession given the direction of the static field $B_0$. For low excitation powers one can neglect the component that rotates opposite to the spin precession, since it can be thought of as being off-resonance by $2\nu_{\rm L}$. This is known as the rotating wave approximation (RWA), valid when $B_1 \ll B_0$.

In the context of quantum computation, the quest for fast and high-fidelity quantum gate operations naturally leads to the tendency to increase the amplitude of $B_1$, eventually up to the point where the RWA breaks down. This regime is difficult to achieve with electron spins, essentially because of the difficulty to produce very large oscillating magnetic fields at gigahertz frequencies. However, an experiment on a nitrogen-vacancy (NV) spin in diamond has shown that breaking the RWA can lead to interesting regimes of time-optimal quantum control~\cite{Fuchs2009}. This could be achieved with an NV spin because it can be addressed and read out at room temperature, allowing large (of order a watt) microwave powers to be applied to the sample. However, the vast majority of solid-state electron spin qubits require operation at sub-Kelvin temperatures, where such high microwave powers would produce unsustainable heating effects.

In this work we investigate the breakdown of the RWA for a single, dressed electron spin in silicon. Coherent dressing of a quantum two-level system has been demonstrated on a variety of systems, including atoms~\cite{Mollow1969}, self-assembled quantum dots~\cite{Xu2007}, superconducting quantum bits~\cite{Baur2009}, NV centres in diamond~\cite{London2013,Rohr2014,London2014}, and the electron spin in silicon~\cite{Laucht2016}. In the dressed basis the eigenstates of the driven system are the entangled states of the photons and the quantum system, which gives rise to a new quantum bit with a level splitting defined by the electron spin Rabi frequency $\Omega_R = \tfrac{1}{2} \gamma_e B_1$~\cite{Cohen2008}. This level splitting is much smaller than the level splitting of the undressed electron spin $\nu_e (= \nu_{\rm L})$, which allows reaching the strong driving regime with moderate microwave powers~\cite{Saiko2007,Avinadav2014}. Furthermore, the dressed qubit can be driven using frequency modulation (FM) of the MW source~\cite{Laucht2016}, which does not add any heating to that of the MW power used for dressing. With this method, we obtain Rabi frequencies of the dressed qubit $\Omega_{R\rho}$ equal to or even slightly larger than its transition frequency $\Omega_R$, which also implies that the dressed spin can be controlled equally fast as the bare electron spin. We observe clear deviations from normal Rabi oscillations and we numerically simulate the time evolution of the states in excellent agreement with the experimental data. The ease with which we can reach this regime and the level of control that we have over the driving fields, make the dressed electron spin a model system for investigating the strong driving regime, where the RWA breaks down and Bloch-Siegert shifts become important~\cite{Bloch1940,Saiko2007,FornDiaz2010}. The development of deterministic quantum control beyond the RWA can have a significant advantage for quantum computation by allowing much faster gate operations to be performed~\cite{Avinadav2014,Deng2015,Song2016}.

The sample investigated here consists of a $^{31}\rm P$ donor that is ion-implanted~\cite{vanDonkelaar2015} into an isotopically purified $^{28}\rm Si$ epilayer~\cite{Itoh2014}. Next to the donor is a single electron transistor that is used for electron spin readout via spin-dependent tunneling~\cite{Morello2010}. A nearby, on-chip, broadband microwave antenna is used to create an oscillating magnetic field for spin control~\cite{Dehollain2013}. The chip is bonded to a printed circuit board inside a copper enclosure and mounted to the coldfinger of an Oxford dilution refrigerator to obtain an electron temperature of $\sim100$~mK. An external magnetic field with $B_0=1.55$~T is applied using a superconducting magnet. Sample and setup are identical to the ones described in Refs.~\onlinecite{Muhonen2014,Laucht2015,Dehollain2015,Laucht2016}, and we refer to those publications for more details about device fabrication and methods.

\begin{figure}[!t]
\begin{center}
\includegraphics[width=1\columnwidth]{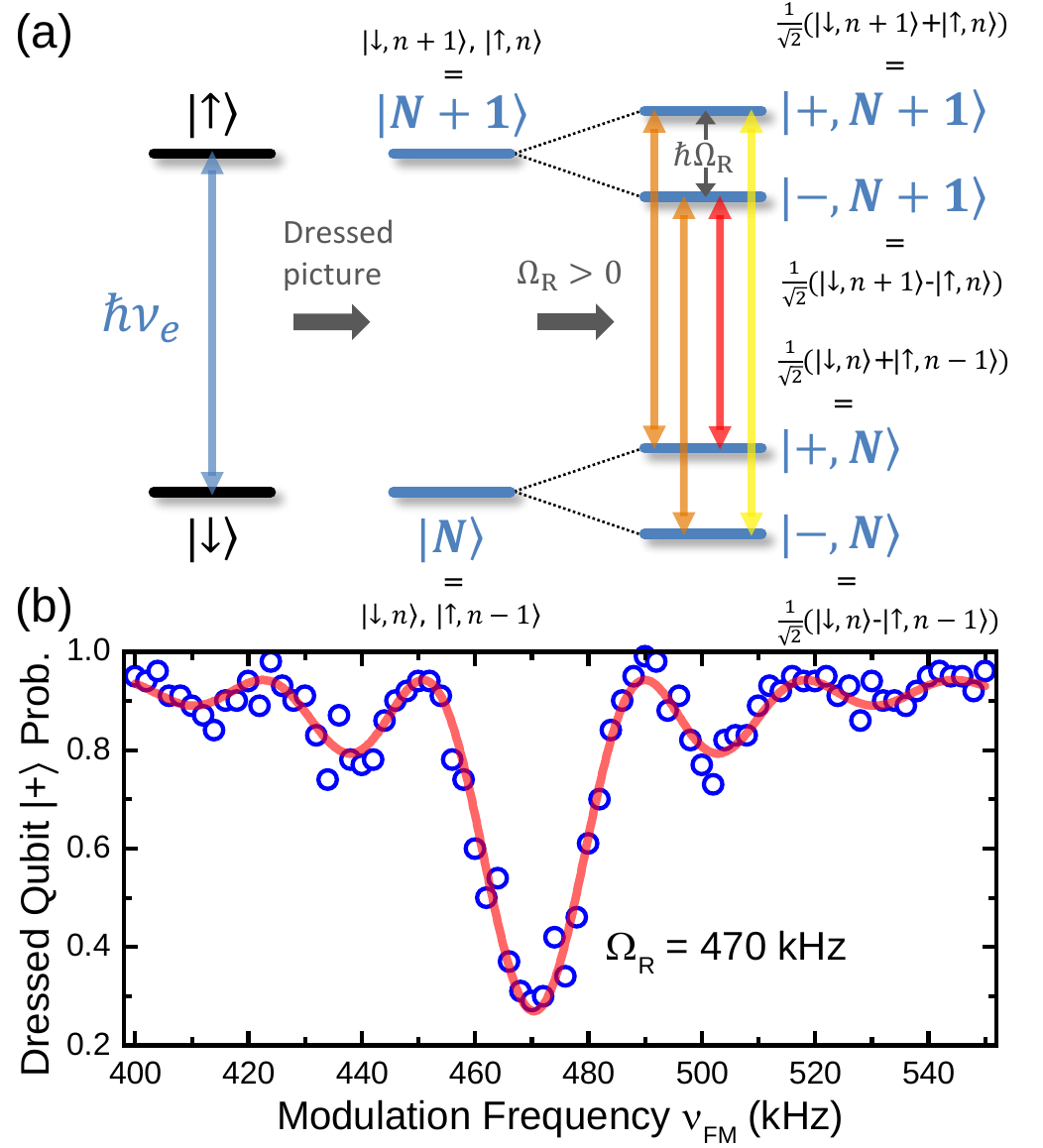}
\caption{\label{figure01} \textbf{The dressed electron spin system.}
(a) Energy level diagram of the electron spin subsystem in the spin picture and the dressed picture.
(b) Dressed qubit Rabi spectrum obtained via frequency modulation resonance.
}
\end{center}
\end{figure}

The spin Hamiltonian of the $^{31}\rm P$ donor in the lab frame is given by
\begin{equation}
\label{Hami0Lab}
H_{0}^{\rm lab} = \gamma_e B_0 S_z + \gamma_n B_0 I_z + A(S_x I_x + S_y I_y + S_z I_z),
\end{equation}
where $S$ and $I$ are the Pauli matrices for the electron spin and the nuclear spin, respectively. The gyromagnetic ratios of electron and nucleus are given by $\gamma_e = 27.97$~GHz/T and $\gamma_n = -17.23$~MHz/T, and the hyperfine coupling between electron and nucleus is $A=96.9$~MHz for this particular donor~\cite{Muhonen2014,Laucht2015}. We can add a resonant drive with 
\begin{equation}
\label{Hami1Lab}
H_{1}^{\rm lab} = B_1 \cos(2\pi\nu_{\rm MW}t) (\gamma_e S_x + \gamma_n I_x),
\end{equation}
where $\nu_{\rm MW}$ is the frequency of the oscillating magnetic field and $B_1$ is its amplitude. As $B_1\ll B_0$ we can make use of the RWA for the oscillating $B_1$ and write the full Hamiltonian in the rotating frame of the MW source 
\begin{equation}
\label{HamiFullRot}
\begin{array}{lll}
H^{\rm rot} &=& (\gamma_e B_0 - \nu_{\rm MW}) S_z + (\gamma_n B_0 - \nu_{\rm MW}) I_z \\
&& + A(S_x I_x + S_y I_y + S_z I_z) \\
&& + \tfrac{1}{2} \gamma_e B_1 S_x + \tfrac{1}{2} \gamma_n B_1 I_x.
\end{array}
\end{equation}
When neglecting the nucleus, the simplified Hamiltonian in the rotating frame is given by
\begin{equation}
\label{HamiRot}
H^{\rm rot} = \Delta\nu S_z + \Omega_R S_x,
\end{equation}
where $\Delta\nu = \nu_e - \nu_{\rm MW}$ is the detuning between the electron spin transition frequency $\nu_e=\gamma_e B_0 + A/2$ for the $\ket{\downarrow}\leftrightarrow\ket{\uparrow}$ transition when the nucleus is in the $\ket{\Uparrow}$ state and the frequency of the microwave source $\nu_{\rm MW}$, and $\Omega_R = \tfrac{1}{2} \gamma_e B_1$ is the Rabi frequency of the electron spin. While resonantly driven at $\Delta\nu=0$, the spin is quantized along the $x$-axis in the rotating frame, with eigenstates $\ket{+}=\tfrac{1}{2}(\ket{\uparrow} + \ket{\downarrow})$ and $\ket{-}=\tfrac{1}{2}(\ket{\uparrow} - \ket{\downarrow})$. These eigenstates correspond to the dressed states $|{+,N\rangle}=\tfrac{1}{\sqrt{2}}(|{\downarrow,n\rangle}+|{\uparrow,n-1\rangle})$ and $|{-,N\rangle}=\tfrac{1}{\sqrt{2}}(|{\downarrow,n\rangle}-|{\uparrow,n-1\rangle})$, where $n$ is the number of resonant photons in the driving field and $N$ is the total number of excitations (i.e. photons plus spin) in the system. In our case, the electron is dressed by a classical driving field, where $n$ is very large and its exact value is unimportant. We can therefore omit $N$ as long as we take into account the magnitude of $B_1$, which defines the electron Rabi frequency $\Omega_R$ and therefore the splitting between the two eigenstates $\ket{-}$ and $\ket{+}$, as indicated in the energy level diagram in Fig.~\ref{figure01}(a). 

\begin{figure*}[!t]
\begin{center}
\includegraphics[width=1\textwidth]{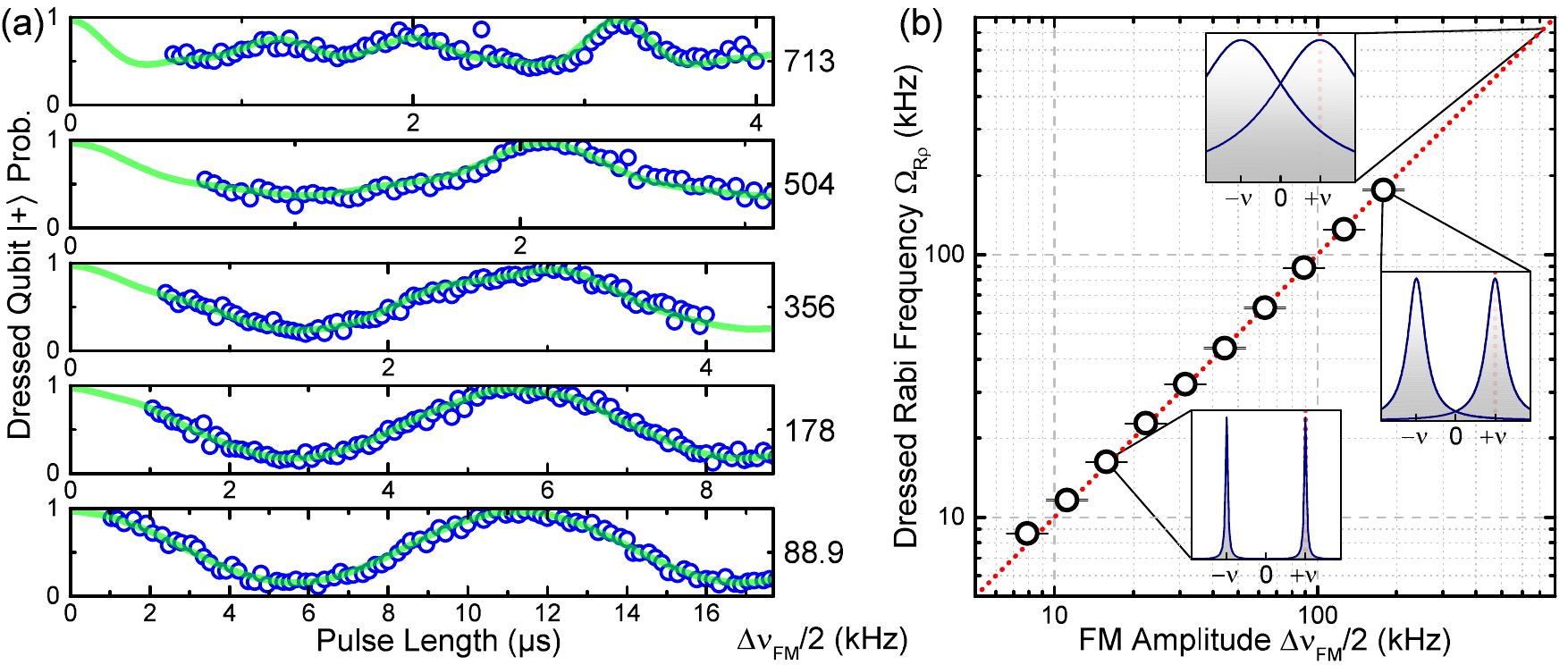}
\caption{\label{figure02} \textbf{Breakdown of the rotating wave approximation.}
(a) Rabi oscillations with FM-drive for different FM amplitudes $\Delta\nu_{\rm FM}$. The solid lines are time evolution simulations that reproduce the dynamics.
(b) Dressed qubit Rabi frequency as a function of $\Delta\nu_{\rm FM}/2$. The insets show the power broadening and overlap of the rotating and counter-rotating waves in the rotating wave approximation.
}
\end{center}
\end{figure*}

We can now rewrite the Hamiltonian $H^{\rm rot}$ in the driven basis with $\ket{+}$, $\ket{-}$ as basis states
\begin{equation}
\label{HamiRho}
H^{\rho} = \Omega_R S_z + \Delta\nu S_x,
\end{equation}
where a modulation of $\Delta\nu$ can be used to control the dressed state. In our experiments, we frequency modulate $\nu_{\rm MW}$ with frequency $\nu_{\rm FM}=\Omega_{R}$ to resonantly drive the dressed spin and coherently control it~\cite{Laucht2016}. Explicitly this means $\Delta\nu(t) = \Delta\nu_{\rm FM} \cos(2\pi\nu_{\rm FM} t + \phi)$, where $\Delta\nu_{\rm FM}$ is the modulation amplitude of the microwave drive, and $\phi$ is the phase of the FM modulation. This is fundamentally different to coherent spin control via interference of multiple Landau-Zener transitions, which is another technique based on frequency modulating the driving field~\cite{Zhou2014}. In the experiments, we achieve FM modulation by connecting a radio frequency (RF) source to the FM input of the MW source. This represents an oscillating RF drive, which resonantly drives the dressed spin with Rabi frequency $\Omega_{R\rho} = \tfrac{1}{2}\Delta\nu_{\rm FM}$, only limited by the maximum modulation amplitude $\Delta\nu_{\rm FM}$ the MW source can achieve. 

We start with an electron initialized in the $\ket{\downarrow}$ state. A MW $\pi/2$ pulse along $-y$ rotates it to the $x$-axis of the Bloch sphere. The phase of the MW source is then switched to $+x$, to keep the electron spin locked along the $x$-axis in the dressed $\ket{+}$ state. This constitutes the initialization of the dressed qubit and its state can now be controlled with a pulse of resonant FM modulation as described above. Fig.~\ref{figure01}(b) shows a spectrum recorded by scanning the modulation frequency $\nu_{\rm FM}$ of the FM pulse over $\Omega_{R}$ ($\Delta\nu_{\rm FM}/2 = 16$~kHz, $T_{\rm pulse}=40$~$\mu$s). Here, the blue circles are the experimental result, while the red line is a fit to Rabi's formula that indicates a coherent rotation of the dressed qubit of $1.28\pi$ in resonance.

We can now record Rabi oscillations of the dressed spin, when we drive it resonantly with $\nu_{\rm FM}=\Omega_{R}$. The use of FM modulation allows us to make $\Omega_{R\rho}$ comparable to, or even larger than the level splitting $\Omega_{R}$ and operate in the regime where the oscillating field cannot be approximated by a rotating field anymore, as the counter-rotating field starts to affect the qubit as well. In Fig.~\ref{figure02}(a) we plot Rabi oscillations obtained for high FM amplitudes from $\Delta\nu_{\rm FM}/2=88.9$~kHz for the bottom to $\Delta\nu_{\rm FM}/2=713$~kHz for the top panel. The blue circles correspond to experimental data, while the green lines are time-evolution simulations (described below). Fig.~\ref{figure02}(b) shows the dressed qubit Rabi frequencies $\Omega_{R\rho}$ extracted from fitting sinusoids to the low FM amplitude experiments (data not shown), following the expected trend. The insets demonstrate the power-broadened excitation linewidths [envelopes from Rabi's formula given by $\Omega_{R\rho}^2/(\Omega_{R\rho}^2+\nu^2)$] for two excitation frequencies at $+\nu$ and $-\nu$. For low $\Delta\nu_{\rm FM}/2<100$~kHz the excitation linewidths are clearly separated, but for higher $\Delta\nu_{\rm FM}/2>100$~kHz the excitation at $-\nu$ starts to affect the qubit at $+\nu$. This is the point where the rotating wave approximation breaks down. The effect of this can be observed in the Rabi oscillations in Fig.~\ref{figure02}(a). For $\Delta\nu_{\rm FM}/2>100$~kHz the Rabi oscillations start to deviate from simple sinusoids~\cite{Fuchs2009}. Although the curve at $\Delta\nu_{\rm FM}/2=713$~kHz seems to follow some unpredictable oscillations, this behaviour is well described by theory. When we model the time-evolution of this strongly driven, dressed qubit we can reproduce the experimental data with excellent qualitative and quantitative agreement [green solid lines in Fig.~\ref{figure02}(a)].

\begin{figure}[!t]
\begin{center}
\includegraphics[width=1\columnwidth]{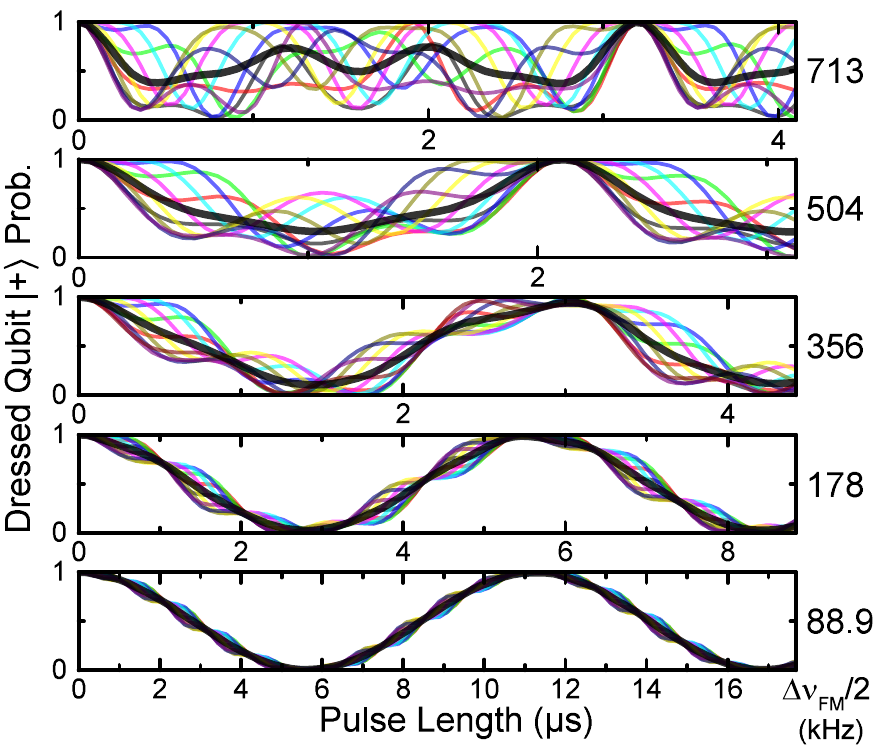}
\caption{\label{figure03} \textbf{Breakdown of the rotating wave approximation.} Time evolution simulations of the dressed qubit orientation when subjected to a driving field at $\nu_e=43.03$~GHz, modulated at $\nu_{\rm FM}=\Omega_R=471$~kHz. The coloured curves were calculated for different phases (0, $\tfrac{1}{10}\pi$, $\tfrac{2}{10}\pi$, ...) of the FM modulation. The thick black curve is the average value of the coloured curves.
}
\end{center}
\end{figure}

For the time-evolution simulations, we use the full Hamiltonian in the rotating frame of the MW drive Eq.~\ref{HamiFullRot}, which describes the effect of $B_1$ immaculately. Here, the level splitting of the dressed qubit is equal to the electron spin Rabi frequency $\Omega_R \approx \tfrac{1}{2} \gamma_e B_1$. We then add the time-dependent Hamiltonian that takes care of driving the dressed qubit via frequency modulating the microwave frequency $\nu_{\rm MW}$ at frequency $\nu_{\rm FM}=\Omega_{R}$ without any approximations:
\begin{equation}
\label{Hami2}
H_2(t) = \Delta\nu_{\rm FM} \cos(2\pi\nu_{\rm FM} t + \phi) (S_z + I_z).
\end{equation}
The phase factor $\phi$ will become important later, and is necessary to explain the observed results. We finally use $H^{\rm rot}$ and $H_2(t)$ to calculate the time evolution operator for discrete time steps $dt=0.1$~ns
\begin{equation}
\label{TOpe}
U_{dt}(t) = e^{-i2\pi (H^{rot} + H_2(t)) dt},
\end{equation}
and use that to step-wise evolve our initial state in time
\begin{equation}
\label{Psi}
\Psi(t) = U_{dt}(t) \Psi(t-dt).
\end{equation}

In Fig.~\ref{figure03} we plot the results of these simulations for different $\Delta\nu_{\rm FM}/2$ from $88.9$~kHz at the bottom to $713$~kHz at the top, corresponding to the data presented in Fig.~\ref{figure02}(a). The different coloured curves were calculated for different phases $\phi$ of the FM modulation during the applied driving pulse, and their deviation from a perfect sinusoid can be understood as the effect of the counter-rotating wave on the dressed qubit transition. This effect gets stronger as the higher driving amplitudes $\Delta\nu_{\rm FM}/2$ power-broaden the excitation spectrum [see also Fig.~\ref{figure02}(b) insets]. Here, $\phi$ defines the phase relation between the resonant, rotating wave and the strongly off-resonant, counter-rotating wave and how the dressed qubit reacts to being subjected to both of them simultaneously. For some phases $\phi$ the rotation of the dressed qubit is initially sped up and then slowed down, while for other phases $\phi$ the rotation is first slowed down and then sped up.

In the experiment, the phase of the RF source generating the FM modulation is not synchronized to the gating pulses, resulting in a random $\phi$ of the driving pulse. This means that the experiment essentially averages over all possible phases $\phi$ (thick black lines in Fig.~\ref{figure03}). The phase-averaged time evolution is almost identical to that obtained in other qubit systems~\cite{London2014}, and taking into account the readout and initialization fidelities, which limit the amplitude of the oscillations, it matches the experimental data almost perfectly [see Fig.~\ref{figure02}(a)]. To exploit the contribution of the counter-rotating wave and obtain fast and deterministic qubit rotations, driving pulses would need to have a fixed phase $\phi$. This can be achieved by synthesizing the driving pulses directly using ultra-fast arbitrary waveform generators~\cite{Fuchs2009,Deng2015,Kim2015}, which nowadays can reach sampling speeds up to $100$~GS/s. On the other hand, the fast oscillatory terms can be suppressed with shaped pulses with slow turn-on and turn-off times that are adiabatic in the Floquet picture~\cite{Drese1999,Deng2015}.

In conclusion, we have presented experiments that show the time-evolution dynamics of a strongly-driven, dressed electron spin in silicon. When the driving strength becomes comparable to the level splitting of the dressed states, we observe a clear deviation from sinusoidal Rabi oscillations. This deviation originates from the effect of the counter-rotating wave on the dressed states and marks the breakdown of the rotating wave approximation. Supporting theory corroborates the measurements and shows that the time evolution is nonetheless deterministic. Here, the use of the dressed electron spin is crucial to the success of the experiment as it allows us to reach the strong driving regime with experimental ease, using driving powers compatible with cryogenic operation. Thus, the dressed electron spin constitutes a model system to study the strong driving regime for spin qubits, and the prospect of attaining fast, time-optimal quantum gate operations.

\begin{acknowledgments}
This research was funded by the Australian Research Council Centre of Excellence for Quantum Computation and Communication Technology (project number CE110001027) and the US Army Research Office (W911NF-13-1-0024). We acknowledge support from the Australian National Fabrication Facility, and from the laboratory of Prof. Robert Elliman at the Australian National University for the ion implantation facilities. The work at Keio has been supported by JSPS KAKEN (S) and Core-to-Core Program.
\end{acknowledgments}

\bibliography{Papers}
\end{document}